\documentclass[%
reprint,
superscriptaddress,
 amsmath,amssymb,
 aps,
]{revtex4-2}

\usepackage{graphicx}%
\usepackage{multirow}%
\usepackage{amsmath,amssymb,amsfonts}%
\usepackage{amsthm}%
\usepackage{mathrsfs}%
\usepackage{xcolor}%
\usepackage{textcomp}%
\usepackage{booktabs}%
\usepackage{listings}%
\usepackage{indentfirst}
\usepackage{float}
\usepackage{hyperref}
\hypersetup{
    colorlinks=true,
    linkcolor=blue,
    filecolor=blue,      
    urlcolor=blue,
    citecolor=blue
    }

\newcommand{\ket}[1]{\left|{#1}\right\rangle}

\usepackage{titlesec}
\titleformat*{\section}{\centering\footnotesize\bfseries\uppercase}

\usepackage[paperwidth=210mm,paperheight=297mm,centering,hmargin=1.7cm,vmargin=1.5cm]{geometry}

\begin{document}
\footnotetext{These authors contributed equally to this work.}
\title{\large \textbf{Quantum Gas Microscopy of Fermions in the Continuum}}
\author{Tim de Jongh}
\thanks{These authors contributed equally to this work.}
\affiliation{Laboratoire Kastler Brossel, ENS-Universit\'{e} PSL, CNRS, Sorbonne Universit\'{e}, Coll\`{e}ge de France, 24 rue Lhomond, 75005, Paris, France}
\author{Joris Verstraten}
\thanks{These authors contributed equally to this work.}
\affiliation{Laboratoire Kastler Brossel, ENS-Universit\'{e} PSL, CNRS, Sorbonne Universit\'{e}, Coll\`{e}ge de France, 24 rue Lhomond, 75005, Paris, France}
\author{Maxime Dixmerias}
\affiliation{Laboratoire Kastler Brossel, ENS-Universit\'{e} PSL, CNRS, Sorbonne Universit\'{e}, Coll\`{e}ge de France, 24 rue Lhomond, 75005, Paris, France}
\author{Cyprien Daix}
\affiliation{Laboratoire Kastler Brossel, ENS-Universit\'{e} PSL, CNRS, Sorbonne Universit\'{e}, Coll\`{e}ge de France, 24 rue Lhomond, 75005, Paris, France}
\author{Bruno Peaudecerf}
\affiliation{Laboratoire Collisions Agr\'egats R\'eactivit\'e, UMR 5589, FERMI, UT3, Universit\'e de Toulouse, CNRS, 118 Route de Narbonne, 31062, Toulouse CEDEX 09, France}
\author{Tarik Yefsah}
\affiliation{Laboratoire Kastler Brossel, ENS-Universit\'{e} PSL, CNRS, Sorbonne Universit\'{e}, Coll\`{e}ge de France, 24 rue Lhomond, 75005, Paris, France}

\date{November 7, 2024}

\begin{abstract}
 \quad Microscopically probing quantum many-body systems by resolving their constituent particles is essential for understanding quantum matter. In most physical systems, distinguishing individual particles, such as electrons in solids, or neutrons and quarks in neutron stars, is impossible. Atom-based quantum simulators offer a unique platform that enables the imaging of each particle in a many-body system. Until now, however, this capability has been limited to quantum systems in discretized space such as optical lattices and tweezers, where spatial degrees of freedom are quantized. Here, we introduce a novel method for imaging atomic quantum many-body systems in the continuum, allowing for in situ resolution of every particle. We demonstrate the capabilities of our approach on a two-dimensional atomic Fermi gas. We probe the density correlation functions, resolving their full spatial functional form, and reveal the shape of the Fermi hole arising from Pauli exclusion as a function of temperature. Our method opens the door to probing strongly-correlated quantum gases in the continuum with unprecedented spatial resolution, providing in situ access to spatially resolved correlation functions of arbitrarily high order across the entire system.
\end{abstract}

\maketitle

Acquiring quantitative knowledge on the microscopic properties of correlated quantum systems is crucial to their deep understanding. Atom- and molecule-based quantum simulators provide a unique approach to pursue this goal, with a high level of control and ever improving imaging capabilities, which in many cases allow resolving each constituent of the system \cite{gross2017, gross2021, kaufman2021, daley2022}. For instance, ultracold atoms and molecules in optical lattices allow to explore Hubbard models, critically important in the context of high-$T_{\rm c}$ superconductivity \cite{jaksch2005, bakr2009, sherson2010, danzl2010, yan2013, gross2017, gross2021, christakis2023, cornish2024}. Tweezer-trapped few-particle systems have enabled studying the build-up of complexity and many-body effects in mesoscopic fermionic clusters, directly relevant to the physics of atomic nuclei \cite{wenz2013, bergschneider2018, bayha2020, holten2021, holten2022, brandstetter2023, lunt2024}. Tweezer arrays of Rydberg atoms and dipolar molecules offer the ability to explore many-body states with long-range interactions and pristine internal state control, allowing to investigate a wide class of spin models \cite{labuhn2016, bernien2017, kaufman2021, cairncross2021, jenkins2022, holland2023}. All of these platforms feature single-particle (atom or molecule) imaging giving direct access to the microscopic properties of the system.

Another important class of quantum gas experiments addresses quantum matter in the bulk, that is, with large ensemble of particles in homogeneous or weakly varying traps, and have had great success in studying several key paradigms of many-body and statistical physics. Prominent examples include studies of fermionic superfluidity in the BEC-BCS crossover \cite{zwierlein2005, zwierlein2006, ku2012, zwerger2012, zwierlein2013}, Berezinksii-Kosterlitz-Thouless topological order \cite{hadzibabic2006, jose2013, desbuquois2012, murthy2015, sobirey2021}, Kibble-Zurek critical dynamics near second-order phase transitions \cite{navon2015, beugnon2017, lee2024}, and supersolid behaviors in Bose gases \cite{li2017, leonard2017, bottcher2019, chomaz2019, recati2023}. Recently the creation of Bose-Einstein condensates of dipolar molecules opened a whole new window of exploration for exotic phases of quantum matter \cite{bigagli2024}. So far, however, the imaging techniques available in these experiments only gave access to average density distributions, and have largely limited their scope to the exploration of global coherence, thermodynamic, or transport properties \cite{zwerger2012,zwierlein2013,beugnon2017,recati2023}. 

%%%
\begin{figure}[t!]
    \centering
    \includegraphics[width=\linewidth]{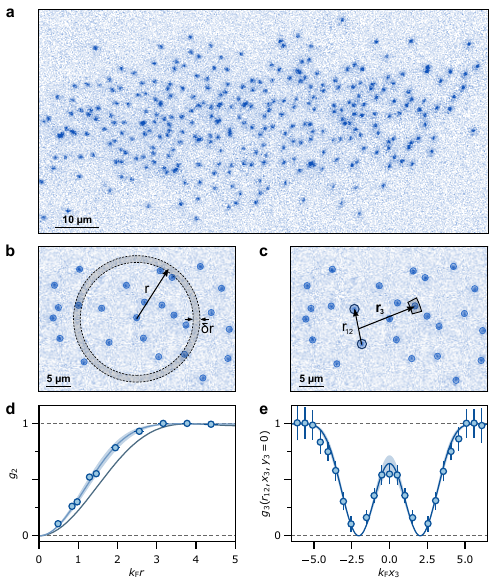}
    \caption{\textbf{Probing \textit{in-situ} correlations in the continuum.} \textbf{(a)} Single-atom image of a non-interacting Fermi gas of $N = 331$ atoms. \textbf{(b)} Extraction of the $g_2$ correlation function from the experimental images. Identified atoms are marked by blue circles. For each atom, we count the number of surrounding atoms at a distance between $r$ and $r+\delta r$, as indicated by the dashed circles. \textbf{(c)} Extraction of the $g_3$ correlation function. We identify all pairs with certain distance $r_{12}$. For each of these pairs, we register the position $\mathbf{r}_3 =(x_3, y_3)$ of each surrounding atom, relative to the center of mass of the pair. \textbf{(d)} Example of measured density-density correlation function $g_2(r)$ (blue circles), zero temperature prediction (grey line), and theoretical fit yielding $T/T_{\rm F} = 0.47(7)$ (blue line, with shaded area indicating uncertainty). \textbf{(e)} Experimental $g_3$ for $r_{12} = 4.1k_{\rm F}^{-1}$ with the third atom positioned along the interparticle axis of the pair ($\mathbf{r}_{12}$), and theory at $T/T_{\rm F} = 0.47(7)$ (solid line and shaded area). Error bars show the standard error of the mean.}
    \label{fig:imgs_and_corrs}
\end{figure}

Here, we introduce quantum gas microscopy in the continuum. We apply it to a two-dimensional (2D) Fermi gas where we spatially resolve each atom in the system in situ (see Fig.\,\ref{fig:imgs_and_corrs}). We measure the full functional form of the two- and three-point spatial density correlation functions, revealing the shape of the Fermi hole as well as its temperature dependence. We study the system in the non-interacting regime, which allows us to directly compare the measured correlation functions to theoretical predictions, finding excellent agreement. Furthermore, we experimentally demonstrate the validity of Wick's theorem \cite{wick1950} for our samples, by relating the measured two- and three-point correlation functions with a high-degree of precision, thus obtaining a stringent validation of our imaging method. Finally, we depart from the pure 2D case by allowing fermions to populate quantum states in the third direction of motion. We show that correlation functions offer a pristine way to characterize quasi-2D Fermi gases, as they provide a quantitative and reliable access to the populations in the transverse motional levels. These measurements represent the first \textit{in-situ} and atom-resolved spatial correlation measurements of a bulk gas. The general approach introduced here is readily applicable to probe strongly-interacting quantum gases. 

\begin{figure*}[t!]
    \centering
    \includegraphics[width=\textwidth]{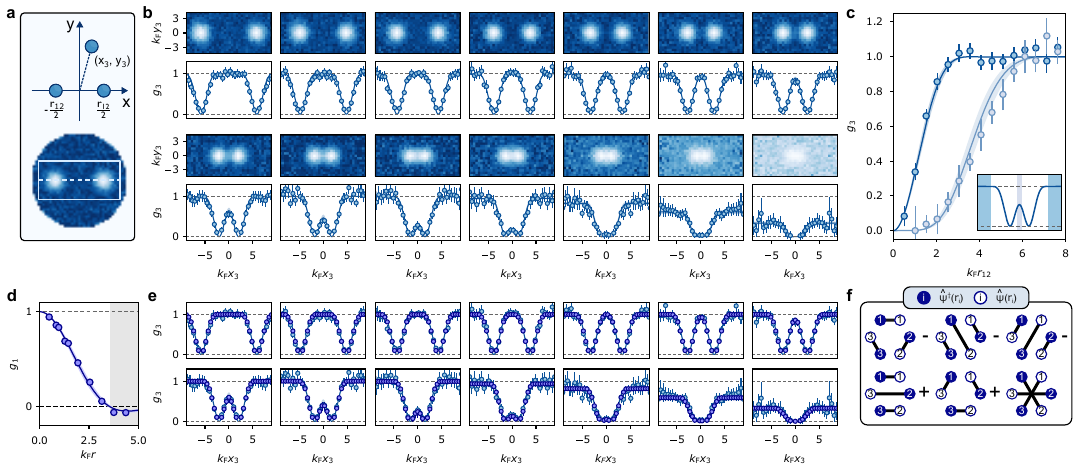}
    \caption{\textbf{Three-body correlation functions and Wick analysis.} \textbf{(a)} Schematic showing the coordinate system definitions (top) and the region of $g_3$ plotted in the sub-panels of panel \textbf{(b)} (bottom). \textbf{(b)} Density plots and central cuts of the $g_3$ correlation function for a non-interacting Fermi gas. Each sub-panel consists of a 2D density plot of $g_3$ and a central cut ($y_3 = 0$, data points) similar to Fig.\,1\textbf{(e)}. For the respective panels we have (left to right, top to bottom): $k_{\rm F}r_{12} = 11.7, 9.7, 8.7, 7.6, 6.6, 5.6, 5.1, 4.1, 3.6, 3.1, 2.6, 2.0, 1.5$ and $1.0$. \textbf{(c)} Average $g_3$-values as a function of $k_{\rm F}r_{12}$ for two limiting cases, as illustrated by the inset. The dark blue data points show $\lim_{r_3\rightarrow\infty}g_3(r_{12},x_3,y_3)$. The light blue data points are taken from the center of each trace in panel \textbf{(b)}, where $x_3 = y_3 \approx 0$. Solid lines in panels \textbf{(b)} and \textbf{(c)} show theory at the temperature obtained from the $g_2$ measurement, with shaded areas indicating uncertainties. \textbf{(d)} Coherence function $g_1(r)$ extracted from the $g_2-$function through Wick's theorem. \textbf{(e)} Comparison between the directly extracted $g_3$ correlation function from panel \textbf{(b)} (blue data points) and the $g_3-$function obtained through application of Wick's theorem on the extracted $g_1$ from panel \textbf{(d)} (purple data points). \textbf{(f)} Visual representation of Wick's theorem applied to the $g_3$ correlation function. Blue (white) circles represent creation (annihilation) operators. Black lines represent the different contractions that allow $g_3$ to be expressed in terms of $g_1$ correlation functions.}
    \label{fig:g3_data}
\end{figure*}

\subsection*{Challenges}
Quantum gas microscopy was initially developed in the context of Hubbard physics \cite{bakr2009,sherson2010} and has so far been devoted to the study of lattice and spin-chain systems \cite{gross2017,gross2021}, where atoms initially evolve in a discretized space and can tunnel from site to site. To image the system, atoms are first pinned by ramping up the lattice depth to a value preventing any tunneling according to a simple and well defined adiabaticity criterion, and subsequently exposed to fluorescence light allowing to detect each atom. Here, in contrast, we are interested in pinning the atoms of many-body systems that initially evolve in continuous space, whose projection dynamics is far more complex due to the absence of an initial energy gap and has not been studied to date. A first crucial challenge is therefore to ensure that the pinning of the many-body wave function preserves the collective information prior to pinning.

A second difficulty stems from light-assisted collisions that occur during imaging when two atoms occupy the same lattice site, such that quantum gas microscopy only gives access to the parity of the occupation number. In the study of Bose- or Fermi-Hubbard systems, the occupation is typically of one or two atoms per lattice site such that this \text{parity projection} can be mitigated \cite{gross2017,gross2021}. In contrast, the high densities typically used in bulk systems correspond to having tens to hundreds of atoms per lattice site, such that parity projection would be crippling for most quantitative measurements. This detrimental effect can be circumvented by working with extremely dilute clouds, with about two orders of magnitude lower densities. The associated challenge is to prepare samples at accordingly lower temperatures in order to reach the deep quantum degenerate regime. In practice, this requires temperatures in the range of 1--20\,nK, which are below or at the lowest end of the temperatures typically reached in bulk quantum gases. 

In this work, we tackle these two challenges and perform quantum gas microscopy in the continuum for the first time, which we describe in the following.

\subsection*{Preparation and Single-Atom Imaging of Ultra-dilute Fermi Gases}
Our experiment starts with a single-component non-interacting Fermi gas of $^6$Li atoms confined in a plane using a `light sheet', a highly oblate optical dipole potential providing strong confinement in the vertical $z$--direction. In the $xy$-plane, the light sheet provides a shallow Gaussian-like trap, which can be approximated by a harmonic potential near the trap center. The corresponding trapping frequencies are $\omega_x/2\pi\approx 30 $\,Hz, $\omega_y/2\pi\approx 80$\,Hz and $\omega_z/2\pi\approx1.1$\,kHz. In-plane density variations near the trap center are nonethless small enough to extract homogeneous quantities, as is typically done in bulk experiments in the framework of the local density approximation \cite{brack2003,castin2007}. Having prepared the system with a given atom number ranging from $N = 45(7)$ to $325(21)$, we suddenly pin the atoms by ramping on a deep optical lattice in the $xy$-plane and apply Raman side-band laser cooling, which reduces the motional energy of each atom in its lattice well and simultaneously drives their fluorescence. The scattered photons are collected through a high-resolution objective and projected onto a CCD camera. 

The pinning phase, where atoms initially evolving in the continuum are projected onto the wells of the lattice, is crucial in our experiment. According to our previous work on single-atom wavepackets \cite{verstraten2024}, reliable pinning requires the lattice ramp-on time $\tau$ to satisfy the inequality $\omega^{-1}\ll\tau\lesssim a_{\rm L}/v$, where $\omega$ is the lattice well frequency at the end of the ramp, $a_{\rm L}$ is the lattice spacing and $v$ the characteristic velocity of the system. While there is no demonstration that this criterion is sufficient at the many-body level, it is reasonable to consider it as a minimal requirement. For the Fermi gases considered here, the relevant scale is set by the Fermi velocity $v_{\rm F}$, spanning from $7$ to $14$ mm/s. Applying our criterion, we set the lattice ramp-time to $\tau=10\,\mu$s (see \cite{verstraten2024}).

In Fig.\,\ref{fig:imgs_and_corrs}a, we show a typical experimental image of a non-interacting single-component Fermi gas obtained with our quantum gas microscope. The cloud shown here contains $N=331$ atoms and is one of our densest clouds, with an inter-particle distance on the order of five times the lattice spacing. In this regime of diluteness, density correlation functions are readily accessible as depicted in Fig.\,\ref{fig:imgs_and_corrs}b and Fig.\,\ref{fig:imgs_and_corrs}c. Since we are interested in density correlations of the homogeneous Fermi gas, we perform the analysis on a central region of the cloud where the local density $n(\mathbf{r})$ is constant within 7\% and the local average Fermi wave vector $k_{\rm F}(\mathbf{r})= \sqrt{4\pi n(\mathbf{r})}$ is constant within 3.5\% (see \footnote{Additional details can be found in the Supplementary Materials}). 

\subsection*{In-situ Correlation Measurements}

While our images allow to access density correlation functions up to the highest order set by the particle number, we focus here on extracting the two- and three-point density correlations. Their reduced forms respectively read:

\begin{equation}
g_2(\mathbf{r}_1,\mathbf{r}_2) =\frac{\langle \psi^{\dagger}(\mathbf{r}_2)\psi^{\dagger}(\mathbf{r}_1)\psi(\mathbf{r}_1)\psi(\mathbf{r}_2)\rangle}{n^2}, 
\label{eq:g2}
\end{equation} and

\begin{equation}
g_3(\mathbf{r}_1,\mathbf{r}_2,\mathbf{r}_3)= \frac{\langle \psi^{\dagger}(\mathbf{r}_3)\psi^{\dagger}(\mathbf{r}_2)\psi^{\dagger}(\mathbf{r}_1)\psi(\mathbf{r}_1)\psi(\mathbf{r}_2)\psi(\mathbf{r}_3)\rangle}{n^3}, 
\end{equation}
with $\psi^{\dagger}(\mathbf{r}_i)$ and $\psi(\mathbf{r}_i)$ the fermionic field operators, $\mathbf{r}_i$ the position of $i^{\rm th}$ atom, and $n$ the local average density. In an isotropic homogeneous system at a given temperature, $g_2$ only depends on $r=|\mathbf{r}_1-\mathbf{r}_2|$ such that it can be expressed as a one-dimensional function, and $g_3$ only depends on the absolute distance $r_{12}=|\mathbf{r}_1-\mathbf{r}_2|$ between two fermions labeled as 1 and 2, and the position of the third atom $\mathbf{r}_3$.  

We start our analysis with Fermi gases that were prepared with the lowest atom numbers $N=45(7)$. Near the center of the cloud, the average density is $n=0.04\,\mu$m$^{-2}$, the average inter-particle spacing $d=1/\sqrt{n}=5.1\,\mu$m, corresponding to an inverse Fermi wave vector $k_{\rm F}^{-1}=1.44\,\mu$m, and a Fermi temperature $T_{\rm F}=19.1(1)$\,nK, defined as $T_{\rm F}= E_{\rm F}/k_{\rm B} = \hbar^2 k_{\rm F}^2/(2 m k_{\rm B})$, with $E_{\rm F}$ the Fermi energy, $\hbar$ the reduced Planck constant, $m$ the particle mass and $k_{\rm B}$ the Boltzmann constant.

In Fig.\,\ref{fig:imgs_and_corrs}d, we show the correlation function $g_2(k_{\rm F}r)$ extracted from the analysis of 2500 images, along with a fit to the theoretical prediction at finite temperature (see \cite{Note2}). This measurement alone contains several key pieces of information and demonstrates that the two aforementioned challenges have been overcome. Firstly, we observe a clear Fermi hole, which directly reveals that our clouds are degenerate while being in the regime of diluteness required to eliminate double-occupancies to a high degree. More quantitatively, we obtain $T=9.1(7)$\,nK with an excellent fit to the experimental data, corresponding to $T/T_{\rm F} = 0.47(7)$, using the absolute temperature $T$ as the sole fitting parameter, since $g_2$ only depends on $k_{\rm F}r$ and $T/T_{\rm F}$, with $k_{\rm F}$ and $T_{\rm F}$ directly accessible via the measured density. Finally, the observation of a 100\%-contrast Fermi hole matching the theoretical prediction indicates reliable pinning of the many-body wavefunction. If a fraction of atoms were recaptured at a random distance from their initial location upon pinning, the contrast would be reduced by an amount on the order of that fraction. While fermionic anti-bunching was observed in previous correlation measurements in momentum/time domain \cite{jeltes2007,thomas2024} and in lattice systems \cite{cheuk2016,hartke2020}, the shape of the Fermi hole was never measured, to the best of our knowledge. 

\begin{figure*}[!ht]
    \centering
    \includegraphics[width=\textwidth]{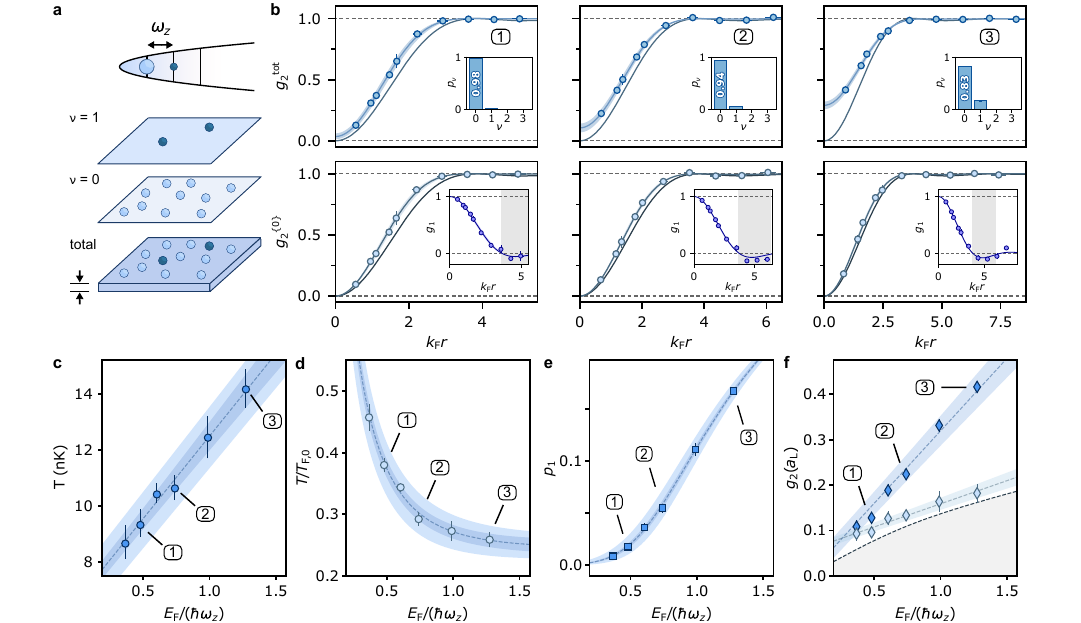}
    \caption{\textbf{Full characterization of a quasi-2D non-interacting Fermi gas.} \textbf{(a)} Schematic representation of the two motional state populations in the vertical harmonic trap potential. Atoms in the $\nu = 0$ and 1 states are represented by the light and dark blue particles, respectively. \textbf{(b)} Top row: Experimental $g_2$-correlation functions for $N = 71, 142$ and $325$ atoms (data points)--labeled 1 through 3, respectively--with theoretical fits using $T$ as the sole fitting parameter (blue curves). The shaded regions represent fitting uncertainties. Grey curves show the $g_2$ of a purely two-dimensional gas at $T = 0$. Error bars represent the standard error of the mean. Insets show the distribution of the vertical level population obtained from the fit. Bottom row: $\nu = 0$ contribution to the $g_2$-correlation function for the respective top panels with insets showing the $g_1$-coherence function obtained from a Wick decomposition. \textbf{(c - e)} Temperatures ($T$), reduced temperatures of the $\nu = 0$ contribution ($T/T_{\rm F,0} = T/(T_{\rm F} p_0)$) and $p_1$-values for systems prepared at different Fermi energies ($E_{\rm F}$). Dashed lines are guides to the eye. Dark and light blue shaded areas indicate statistical and systematic uncertainties, respectively. \textbf{(f)} Value of $g_2^\mathrm{tot}$ (dark blue) and $g_2^{\lbrace 0\rbrace}$ (light blue) at a distance of one lattice site ($a_{\rm L}$). The grey dashed line shows the value of $g_2(a_{\rm L})$ in the pure-2D zero-temperature limit. Blue dashed lines and shaded areas serve as guides to the eye.}
    \label{fig:g2_data}
\end{figure*}

From the same set of images, we extract the three-point reduced density correlation function $g_3(k_{\rm F}r_{12}, k_{\rm F}\mathbf{r}_3)$,  by selecting a pair of atoms and evaluating the probability to find a third atom at a given relative coordinate in the $xy-$plane. Applying this procedure to all possible pairs within the central region of the cloud we obtain density plots as shown in Fig.\,\ref{fig:g3_data}a. To visually represent $g_3$, we set the origin of space between two fermions labeled 1 and 2, such that $\mathbf{r}_2=-\mathbf{r}_1=(r_{12}/2)\mathbf{e_x}$ and $\mathbf{r}_3=x_3\mathbf{e_x}+y_3\mathbf{e_y}$.  We show a set of $g_3$ functions in Fig.\,\ref{fig:g3_data}b for different values of the distance $r_{12}$ that are all extracted from the same data set, and which we represent both as density plots and as cuts taken at $y_3 = 0$. In the density plot, the position of the two fixed atoms is evident from the Fermi hole surrounding them. As the distance $r_{12}$ is  reduced from several times to a fraction of $k_{\rm F}^{-1}$, we not only see the Fermi holes progressively merge but also the probability to find a third neighboring atom drastically drops. This observation provides a striking visualization of the Pauli exclusion principle: in an ideal Fermi gas, particles tend to maximize space occupation and the probability of finding two atoms or more within an area of radius $\lesssim k_{\rm F}^{-1}$ is extremely small. 

To test whether the two-point and three-point density correlation functions are mutually consistent and to which extent, we compare the cuts shown in Fig.\,\ref{fig:g3_data}b to the theoretical prediction at the temperature $T=9.1(7)$\,nK, which is the temperature obtained from fitting $g_2$, and find excellent agreement throughout the entire range of $r_{12}$ values. Upon fitting the $g_3$ measurements to the theoretical prediction, leaving the absolute temperature as a free parameter, we find $T=8.7(3)$\,nK.

The ideal Fermi gas Hamiltonian being quadratic in creation and annihilation operators, Wick's theorem predicts that any correlation function can be expressed in terms of the coherence function $g_1(r)= \langle \psi^{\dagger}(0)\psi(r)\rangle/n$. Specifically, $g_2 (r) = 1-g_1(r)^2$ and $g_3(\mathbf{r}_1,\mathbf{r}_2,\mathbf{r}_3) = 1 - g_1(r_{12})^2-g_1(r_{23})^2-g_1(r_{31})^2 + 2g_1(r_{12})g_1(r_{23})g_1(r_{31})$ where $r_{ij}=|\mathbf{r}_j-\mathbf{r}_i|$. Using these relations, we first extract the coherence function $g_1$ using the measured $g_2$ function (see \cite{Note2}), which we show in Fig.\,\ref{fig:g3_data}d, and in turn use $g_1$ to obtain $g_3$ for the various values of $r_{12}$. The resulting traces for $g_3(k_{\rm F}r_{12}, x_3, y_3=0)$ are plotted in Fig.\,\ref{fig:g3_data}e, showing a remarkable agreement with the measured ones. This analysis, which does not rely on any fitting, not only shows that the measured $g_2$- and $g_3$-functions are mutually consistent to a high degree, but also represents a demonstration of the validity of Wick's theorem for our samples.

\subsection*{Correlations in a Quasi-2D Fermi Gas}
In the vertical direction the atomic motion is quantized, with a quantum of vibration corresponding to a temperature scale $T_z \equiv \hbar \omega_z/k_{\rm B} = 52(2)\,$nK. For the preparation discussed above, we ensured a minimal occupation of the excited $z$-level states by keeping $T$ and $T_{\rm F}$ sufficiently smaller than $T_z$. We now deliberately prepare clouds for which higher $z$-levels are occupied, by increasing $T_{\rm F}$ to values on the order of $T_z$, while keeping $T$ low relative to both of these scales. For simplicity, we restrict ourselves to preparing samples where only the first excited $z$-level is significantly occupied and the population in higher $z$-levels is kept well below 1\%. Having atoms in the ground and the first excited $z-$levels is equivalent to having two independent 2D Fermi gases, as depicted in Fig.\,\ref{fig:g2_data}a. The total 2D correlation function can be obtained by summing over the $z-$levels, and reads (see \cite{Note2}):
\begin{equation}
g_2^{\rm tot}(r)= 2p_0p_1 + p^2_0 g^{\lbrace 0\rbrace}_2(r) + p^2_1 g^{\lbrace 1\rbrace}_2(r),
\label{eq:gtot}
\end{equation}
where $p_\nu$ is the fraction of atoms in the $z-$level $\nu$, and $g^{\lbrace \nu\rbrace}_2(r)$ the two-point reduced density correlation within this level. From Eq.\,(\ref{eq:gtot}) it follows that a non-zero population in $\nu=1$ yields a non-zero value of the $g_2$-function at short distance, tending to $g_2^{\rm tot}(0)=2p_0p_1$ for zero distance. The Fermi-hole contrast hence provides an excellent probe of the population on the first excited level. The result Eq.\,(\ref{eq:gtot}) can be generalized to an arbitrary number of populated excited state (see \cite{Note2}), and represents a key new possibility for the exploration of low-dimensional quantum gases, where the determination of transverse state populations is typically non-trivial \cite{hadzibabic2008,tung2010,yefsah2011,yefsah2011a,dedaniloff2021}.

In Fig.\,\ref{fig:g2_data}b (top row), we show the total $g_2$-function obtained for samples with increasing Fermi energy. The data is fitted to Eq.\,(\ref{eq:gtot}) using $T$ as the only free parameter (see \cite{Note2}), showing excellent agreement and clearly demonstrating the reduction of the Fermi hole depth with increasing values of $T_{\rm F}$. With the knowledge of $T$, we obtain not only the populations $p_\nu$, but also both $g^{\lbrace 0\rbrace}_2(r)$ and  $g^{\lbrace 1\rbrace}_2(r)$. The function $g^{\lbrace 0\rbrace}_2(r)$ is shown in the bottom row of Fig. \ref{fig:g2_data}b providing a measure of the temperature dependence of the Fermi hole. The lower panels in Fig. \ref{fig:g2_data} displays all the relevant observables of the system and illustrate the power of our method as a diagnostic tool. Note that the sample with the lowest $E_{\rm F}$ corresponds to the  preparation presented in Figs.\,\ref{fig:imgs_and_corrs} and \ref{fig:g3_data} where we now allow for occupation in the $\nu = 1$ state by fitting the data to Eq.\,(\ref{eq:gtot}). This results in a ground-state population $p_0=99(1)\%$, implying that less than one atom occupies the $\nu=1$ level, and validating the assumption that these samples are well within the 2D regime.

\subsection*{Conclusion and Outlook}

We have demonstrated quantum gas microscopy in the continuum. Key elements include the reliable pinning of the many-body wave function from continuous space, and performing parity-projection-free quantum gas microscopy by creating ultra-dilute degenerate bulk systems. We demonstrated the validity and the power of our imaging method via the measurement of two- and three-body density correlations of two-dimensional Fermi gases, which we found to be in excellent agreement with theoretical predictions and to satisfy Wick's theorem. These measurements represent the first \textit{in-situ} and atom-resolved spatial correlation measurements of a bulk gas, complementing previous work with multi-channel plate detectors in momentum space \cite{jeltes2007,chang2016,thomas2024}. 

Our results extend the applicability of quantum gas microscopy to the realm of many-body physics in the continuum, offering a new set of possibilities for the exploration of strongly-correlated quantum states. Our imaging method can be readily used to probe spin-correlations in the 2D and 3D BEC-BCS crossover \cite{Daix2024}, and in particular in the paradigmatic unitary Fermi gas \cite{zwerger2012,zwierlein2013}. It provides a unique opportunity to facilitate the search for the elusive phases of matter such as the Fulde-Ferell-Larkin-Ovchinikov superfluid state \cite{fulde1964,larkin1965}, or quantum Hall states in rotating atomic gases \cite{holland2023} beyond the few-particle regime \cite{leonard2023,lunt2024}. In combination with matter wave magnification techniques \cite{brandstetter2023}, our approach also makes the observation of crystaline order in 2D dipolar systems possible \cite{buchler2007}.

\section*{Acknowledgments}
We thank Fabrice Gerbier and Martin Zwierlein for discussions, and Jean-Paul Nohra and Sebastian Will for a critical reading of the manuscript. We are grateful to Antoine Heidmann for his crucial support as head of Laboratoire Kastler Brossel. This work has been supported by Agence Nationale de la Recherche (Grant No. ANR-21-CE30-0021), CNRS (Tremplin@INP 2020), and R{\'e}gion Ile-de-France in the framework of DIM SIRTEQ and DIM QuanTiP. 

\textit{Note} -- While writing the manuscript, we became aware of related work on \textit{in situ} studies of correlations in ultracold Fermi and Bose gases \cite{Zwierlein2024,Ketterle2024}. Preprint \cite{brandstetter2024} reported  microscopy of magnified few-fermion systems.

\newpage

\section*{Supplementary Materials}\label{sec:Supplementary}

\subsection*{Preparation and Detection of 2D Ideal Fermi Gases.}
Our experimental setup and the generic sequence is described in \cite{jin2024}. In this work, the final part of the sequence starts with a spin-balanced mixture of the two lowest $^6$Li hyperfine states, denoted $\ket{1}$ and $\ket{2}$, loaded into the light sheet potential described in the main text. The frequency ratios of trapping potential are $\omega_z/\omega_y\approx13.5$ and $\omega_z/\omega_x\approx36$ such that about 200 atoms are required to fill the vibrational ground state in the $z$-direction at $T=0$ .
 
After evaporative cooling at 832\,G, the magnetic field is tuned to $585$\,G in order to perform a radio-frequency transfer of all atoms in state $\ket{2}$ to the third-lowest hyperfine state $\ket3$, using a Landau-Zener sweep. At this magnetic field, the scattering lengths $a_{12}$ and $a_{13}$ between states $\ket{1}$ and $\ket{2}$, and $\ket{1}$ and $\ket{3}$, respectively, are both $\sim 264 a_0$, with $a_0$ the Bohr radius, such that any density-dependent effects on the radio-frequency transfer efficiency are mitigated. The magnetic field is then ramped to $320$\,G where $a_{13} = -950a_0$ and evaporation is continued by lowering the light sheet power for a duration of $\sim$14\,s to a variable power with which we control the final total atom number per spin state. Atoms are held in the light sheet for 2.8\,s for thermalization before its trap depth is adiabatically ramped up to 240\,nK in $80$\,ms, corresponding to $\omega_z = 2\pi \times 1.1(1)$\,kHz. This is followed by a removal of all state $\ket{1}$ atoms using optically resonant light propagating in the $z$-direction, i.e., perpendicularly to the plane of the light sheet. Owing to the diluteness of the cloud and the weak interactions, the removal of state $\ket{1}$ does not affect the state $\ket{3}$ atoms. Finally the magnetic field is ramped to 0\,G in $10$\,ms and after a brief hold time ($10$\,ms) atoms are pinned to record their positions.

Pinning is performed according to the protocol described in \cite{verstraten2024}. The pinning lattice is created through the self-interference of a red-detuned 1064\,nm laser. Using three arms crossing at 120$^\circ$ angles in the $xy$-plane this results in a triangular lattice configuration with a spacing of $a_\mathrm{L} = 709$\,nm \cite{jin2024}. At the maximum laser power used in this work (30\,W), the characteristic frequency of the lattice wells is $\omega=2\pi\times 1.0(1)\,$MHz. With the lattice initially off, we pin the atoms by ramping both the light sheet and the $xy$-lattice to their maximum power in $10$\,$\mu$s and initiating Raman sideband cooling (RSC). RSC is applied as soon as the lattice is ramped on \cite{verstraten2024} and camera exposure is initiated after a 2\,s hold time to reduce background light fluctuations. Resulting images are analyzed through a high-fidelity neural network recognition algorithm to obtain the positions of the individual atoms \cite{verstraten2024}. For each experimental iteration, we capture two images with a 600\,ms exposure time and a 250\,ms hold time between successive exposures, all while maintaining the RSC and lattice beams. This allows us to quantify the fraction of atoms that remains pinned in their lattice sites in real time, exceeding 99.9\%.\\

\subsection*{Extraction of correlation functions.}
We extract correlation functions directly from the detected atomic positions. We obtain two-point reduced correlation functions by creating a histogram of the relative atomic positions in the central region of the cloud where the density is essentially homogeneous, with a standard deviation of $k_{\rm F}(\mathbf{r})$ below 3.5\%. For a given image we relate the occupation $N(\mathbf{r}_1)$ of a lattice site $\mathbf{r}_1$ within the central region of interest ROI A to the occupation $ N(\mathbf{r}_2)$ at all other sites $\mathbf{r}_2$, in a slightly larger region of interest ROI B that includes ROI A. We then extract a histogram of $N(\mathbf{r}_1)N(\mathbf{r}_2)$, arranged per relative distance $\mathbf{r} = \mathbf{r}_2-\mathbf{r}_1$ normalized by the local average occupation number of the respective lattice sites $\langle N(\mathbf{r}_1) \rangle$ and $\langle N(\mathbf{r}_2) \rangle$. The individual occupation numbers $N(\mathbf{r})$ are either 0 or 1 such that summing over all pairs of lattice sites is equivalent to summing over all atomic pairs. Repeating this procedure for all lattice sites in ROI A and averaging over all images, we obtain the two-body density correlation function:
\begin{equation}
g_2(\mathbf{r}) = \biggl< \sum_{\mathbf{r}_1 \in A}\sum_{\mathbf{r}_2 \in B} \frac{\delta_{\mathbf{r}, \mathbf{r}_1 - \mathbf{r}_2}}{\eta} \frac{ N(\mathbf{r}_1)N(\mathbf{r}_2)}{\langle N(\mathbf{r}_1) \rangle \langle N(\mathbf{r}_2) \rangle} \biggr>
\end{equation}
where the sums go over all sites in ROI A and ROI B respectively. The $\delta$-symbol ensures only the pairs at the considered interparticle distance are taken into account and $\eta = (\sum_{\mathbf{r}_1 \in A} 1)$ is a normalization on the number of lattice sites in ROI A. We finally perform azimuthal averaging of the $g_2$-function. For each preparation, we obtain $k_{\rm F}(\mathbf{r}) = \sqrt{4\pi n(\mathbf{r})}$, from the average density $n(\mathbf{r})$ over all images.

We obtain the three-point reduced density correlator $g_3$ by considering all atom pairs $1$ and $2$ at interparticle distance $\mathbf{r}_{12}$ within ROI B that have their center of mass $\mathbf{r}_\mathrm{COM}^{(12)}$ within ROI A. We create a histogram of the distribution of third atoms relative to this pair, again normalizing on the average occupation numbers of the respective lattice sites. For each triplet we take $\mathbf{r}_\mathrm{COM}^{(12)}$ as the origin and rotate the coordinate frame such that the interparticle axis of atoms 1 and 2 lies along the x-axis, as indicated in Fig.\,\ref{fig:g3_data}a of the main text. We repeat this for each image and each value of $r_{12} = |\mathbf{r}_{12}|$ to obtain $g_3(r_{12},x_3, y_3)$.

Due to the large number of ways in which three atoms can be distributed over the triangular lattice, we directly bin the relative positions into a cubic grid, where the dimensions correspond to $r_{12}$, $x_3$ and $y_3$, with bin sizes slightly larger than $a_\mathrm{L}$. To avoid analysis artifacts induced by this binning, we normalize the extracted $g_3$ with a function obtained by performing the same analysis for simulated images of uncorrelated, i.e., randomly distributed, atoms with the same average density distribution $\langle n(\mathbf{r}) \rangle_\mathrm{2D}$ as obtained experimentally. This thus corresponds to the correlations of a homogeneous, classical gas for which $g_3(r_{12},x_3, y_3) = 1$ for all values of $r_{12}, x_3$ and $y_3$.\\

\subsection*{Wick Analysis}
We compare the $g_3(k_{\rm F}r_{12},k_{\rm F}x_3,k_{\rm F}y_3)$ correlation function as directly extracted from the experiment with those obtained from the measurement of $g_2(k_{\rm F}r)$ through a Wick-decomposition. Here, we restrict ourselves to the central cuts ($y_3 = 0$) of the pure-2D data set as shown in Fig.\,\ref{fig:g3_data} of the main text. We compute $g_1(k_{\rm F}r) = \pm\sqrt{1 - g_2 (k_{\rm F}r)}$ setting the sign according to the theoretical value of $\mathrm{sgn}(g_1(k_{\rm F}r))$. This is indicated in Fig.\,\ref{fig:g3_data}c and the insets of Fig.\,\ref{fig:g2_data}b by the grey shaded regions. From this we extract $g_3$ by applying the Wick decomposition as given in the main text.\\

\subsection*{Theoretical Correlation Functions}
\begin{figure}[!h]
    \centering
    \includegraphics[width=\linewidth]{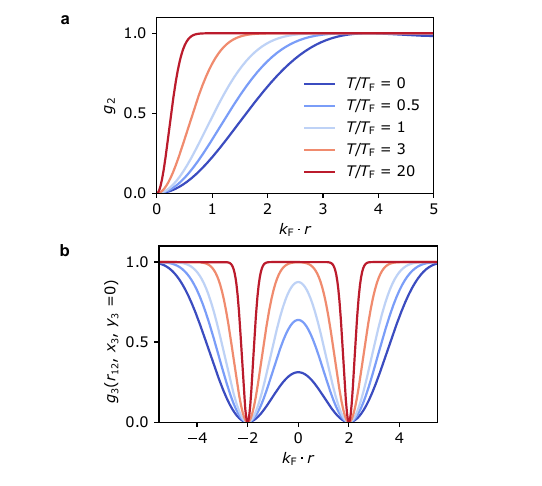}
	\caption{\textbf{Temperature dependence of the correlation functions.} \textbf{(a)} Two-body correlation function at several reduced temperatures. \textbf{(b)} Central slices ($y_3 = 0)$ of the three-body correlation function for $k_{\rm F} r_{12} = 4$ using the same reduced temperatures as in panel (a).}
    \label{fig:theory}
\end{figure}

The theoretical coherence function $g_1(r)$ for a purely two-dimensional non-interacting Fermi gas at finite temperature is computed through a Fourier transform of the Fermi-Dirac distribution. Two- and three-body correlators are then obtained from a Wick decomposition, which yields results that are accurate to better than a few $10^{-3}$ when compared to those obtained from Monte-Carlo calculations. The $g_1$-function  only depends on the interparticle distances $k_{\rm F}r$ and reduced temperatures $T/T_{\rm F}$. Fig.\,\ref{fig:theory} shows computed correlation functions for a purely two-dimensional gas at several values of $T/T_{\rm F}$.

\subsection*{Vertical Harmonic Oscillator Levels}
The trapping potential of the light sheet can be well-approximated by a harmonic oscillator in all three directions, allowing decoupling of the motion along the three different axes. While the trap frequencies in the $xy$-plane are low enough such that the local density approximation holds, the vertical trap frequency is too large to neglect quantization of motion. Occupation of the different motional $z$-levels is determined by both the Pauli exclusion principle and the temperature. For low temperatures and atom numbers only the ground state of the vertical motion is occupied and the non-interacting Fermi gas is purely two-dimensional, such that the theoretically computed correlation functions discussed above can be readily used to fit the temperature of the system. But when the temperature or the total atom number increases, multiple motional states in the $z$-direction become occupied.

Atoms in the different $z$-level are distinguishable such that their total density distribution can be written as a sum of non-interacting 2D layers. For a homogeneous, thermal gas the total density is given by:
\begin{equation}
n = \frac{1}{\lambda^2_\mathrm{T}}\sum_{\nu=0}^\infty{\ln{(1 + e^{\beta\mu_{\nu}}})},
\label{eq:nq2D}
\end{equation}
with $\lambda_\mathrm{T}$ the thermal de Broglie wavelength, $\beta = 1/k_{\rm B}T$ and a $z$-level dependent chemical potential $\mu_\nu = \mu_0 - \nu\hbar\omega_z$, where $\mu_0$ is the global chemical potential. The fraction of atoms $p_\nu$ in $z$-level $\nu$ is then:
\begin{equation}
p_\nu = \frac{\ln{(1 + e^{\beta\mu_{\nu}}})}{\sum_{\nu=0}^\infty{\ln{(1 + e^{\beta\mu_{\nu}}})}}.
\label{eq:pnu}
\end{equation}
The total density correlations of the system $g_2^\mathrm{tot}(r)$ can then be written as:
\begin{align}
g_2^\mathrm{tot}(r) &= \sum_{\nu, \nu'} \frac{\langle \psi_\nu^{\dagger}(\mathbf{r}_2)\psi_{\nu'}^{\dagger}(\mathbf{r}_1)\psi_{\nu'}(\mathbf{r}_1)\psi_\nu(\mathbf{r}_2)\rangle}{n^2} \nonumber \\
 &= \sum_\nu p_\nu^2 g_2^{\lbrace \nu\rbrace}(r) + \left(1 - \sum_\nu p_\nu^2\right),
 \label{eq:g2tot_q2D}
\end{align}
where $g_2^{\lbrace \nu\rbrace}(r)$ is the two-point reduced density correlation within level $\nu$ given by:
\begin{align}
g_2^\mathrm{\lbrace \nu\rbrace}(r) &= \frac{\langle \psi_\nu^{\dagger}(\mathbf{r}_2)\psi_{\nu}^{\dagger}(\mathbf{r}_1)\psi_{\nu}(\mathbf{r}_1)\psi_\nu(\mathbf{r}_2)\rangle}{n_\nu^2} \nonumber \\
& = g_2\left(\sqrt{p_\nu}k_Fr, \frac{T}{p_\nu T_F} \right),
\label{eq:gnu}
\end{align}
where $n_\nu = p_\nu n$.
With $T_{\rm F}$ and $k_{\rm F}$ directly obtained from the density, and $\omega_z$ independently calibrated, the determination of $p_\nu$ only requires the systems temperature $T$ through Eq.\,\eqref{eq:pnu}. The experimentally obtained correlation function $g_2^\mathrm{tot}(r)$ can thus be fitted with $T$ as the sole fitting parameter.

In the present work, only the $\nu = 0$ and 1 levels are populated such that Eq.\,\eqref{eq:g2tot_q2D} reduces to Eq.\,\eqref{eq:gtot} of the main text. We have verified this by fitting the correlation functions using Eq.\,\eqref{eq:g2tot_q2D} while including higher $z$-levels, indeed observing that the occupation of levels with $\nu>1$ remains well below 1\%.

\begin{figure}[!t]
    \centering
    \includegraphics[width=\linewidth]{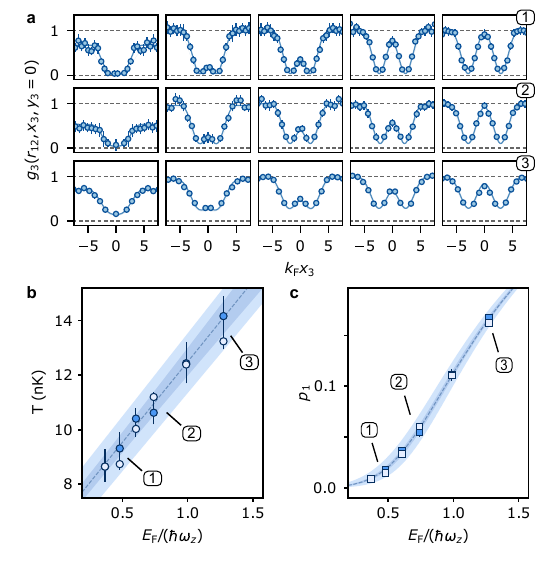}
	\caption{\textbf{Three-body correlations in a quasi-2D non-interacting Fermi gas.} \textbf{(a)} Each row shows central cuts of the $g_3$ reduced density correlations for samples prepared at different Fermi energies $E_{\rm F}$, labeled 1 through 4. Each column corresponds to approximately the same value of $k_{\rm F} r_{12} = $ 1.7, 2.8, 3.6, 4.5 and 5.8. In each row, the solid lines are theoretical predictions resulting from a joint fit of all experimental cuts (25 for each preparation) with the temperature as a single free parameter. \textbf{(b)} Temperatures obtained from theory fits of the $g_2$ data (dark blue) and $g_3$ data (light blue). Dark (light) blue shaded areas show statistical (systematic) errors in the temperature obtained from the fit to the two-body density correlation function, as shown in Fig.\,\ref{fig:g2_data} of the main text. \textbf{(c)} First excited $z$-level populations obtained from the Fermi-Dirac distributions using the temperatures in panel (b) for both the $g_2$ (dark blue) and $g_3$ (light blue) fits, with shaded areas corresponding to those shown in panel (b). The dashed lines and shaded areas serve to guide the eye.}
    \label{fig:g3_q2D}
\end{figure}

\subsection*{Three-body correlation functions in a Quasi-2D Fermi Gas}
In Fig.\,\ref{fig:g3_q2D}a we show the central traces of the experimentally obtained three-point reduced density correlations $g_3(r_{12}, x_3, y_3 = 0)$ obtained from quasi-2D Fermi gases. Each row corresponds to a specific Fermi energy, while each column corresponds to a particular value of $k_{\rm F}r_{12}$. The increase in Fermi energy leads to a higher number of atoms in excited $z$-levels, reducing the contrast of the Fermi hole, which is determined by $2p_0p_1$. This reduction of contrast is clearly observed in each column of Fig.\,\ref{fig:g3_q2D}. For each preparation, we also extract temperatures from a theoretical fit to the sets of $g_3$ traces, taking all $z$-levels into account and again using only $T$ as a fit parameter. Fitted values of $T$ and $p_1$ are shown in Fig.\,\ref{fig:g3_q2D}b and c, and compared to the values obtained from the two-body correlations shown in Fig.\,\ref{fig:g2_data}. Both temperatures agree up to a mean difference of less than 1\,nK. These results illustrate our capability to fully characterize a quasi-2D Fermi gas through its two- and three-body correlations, only needing prior knowledge of the quantization of motion in the vertical direction.\\

\end{document}